# The OAPS solution: a real-time predictive system for flexible PWR operation

Guillaume Dupré[1*], Alain Grossetête[1]

[1]Reactor Process Department, Framatome, 1 Pl. Coupole Jean Millier, Courbevoie, 92400, France
[*]Corresponding author: guillaume.dupre@framatome.com

This paper presents an innovative solution designed to facilitate safe and flexible operation of nuclear power plants. The purpose of this new device, named OAPS system, is to provide optimal strategies (e.g., axial offset control, xenon oscillations mitigation, effluent minimization) and real-time recommendations (e.g., dilution and boration flowrates, turbine power setpoints and variation rates) to help NPP operators perform power variations confidently and efficiently. In fact, just as a GPS navigator optimizes and modifies its planned route according to the current position of the user, the OAPS system regularly updates its recommendations based on the latest plant measurements. To achieve this, the OAPS system relies on a well-established – yet cutting-edge in the nuclear industry – advanced control technique known as model predictive control. The conventional axial offset control strategy of the OAPS system was previously validated on both Framatome's full-scope PWR simulator and EDF's full-scope N4 simulator. In this paper, three new advanced strategies are showcased on an intermediate-complexity PWR simulator developed by Framatome: 1) determination of the fastest feasible power variation rates, 2) accelerated cancellation of axial power oscillations and 3) minimization of water and boron effluents.

## 1. New flexibility challenges for nuclear reactors

### 1.1. From base-load to load-following operation

In order to stabilize grid frequency, electricity consumption and production must be balanced at all times. To achieve load balancing on the generation side, a number of power plants need to be flexible, i.e., capable of adjusting their power output to grid demand [1]. This mode of operation is the most economically attractive, given the low marginal costs of nuclear generation compared to fossil fuels, which have traditionally been used to compensate for daily load fluctuations. Nowadays, however, the base-load operation of NPPs is often challenged by the growing use of renewable energy sources (primarily wind and solar power) and the gradual phase-out of controllable fossil fuel sources [2]. In fact, NPPs are increasingly called upon to balance electricity consumption and production, especially when fossil-fuel power plants are unable to offset excess energy from intermittent renewable sources (e.g., on unexpectedly sunny or windy days). Such situations generally result in low or even negative prices on the electricity market (see Figure 1), prompting NPP operators to rapidly reduce the power output of their units. Conversely, NPP operators strive to quickly return to higher power levels once electricity prices become profitable again. Unfortunately, this is not always easy or possible to achieve, as flexibility can be limited by the core control system of the NPP.

### 1.2. From planned to short-notice power transients

Control room operators must perform several manual actions to ensure the NPP is operated safely and efficiently. Usually, one operator is in charge of the reactor core (the "primary" operator), while the other is in charge of the turbine island (the "secondary" operator). During power variations, the main task of the primary operator is to control the AO of the reactor core by changing the boron concentration of the primary circuit. The secondary operator, on the other hand, has to set the power target and the power variation rate of the turbine according to the grid dispatcher instructions. Under normal circumstances, NPP operators receive grid instructions several hours in advance and can rely on simulation tools to help them prepare for planned power variations [3]. These preparation tools are typically used to calculate the future volumes and flowrates of borated or fresh water that must be injected into the primary circuit to control the AO. However, the results can vary significantly from one transient to another, depending on the amplitude and speed of the power variations as well as the reactor core history and the fuel burnup. Thus, in practice, NPP operators must continuously monitor and adjust the pre-computed recommendations, since the predictions of the preparation tools are never perfect. This demands a high-level of concentration and alertness, even for experienced operators, as inaccurate adjustments can slow down or even cancel power variations. In recent years, the growing share of intermittent renewable energy sources has made the grid less predictable and more complex to manage. As a result, NPP operators are increasingly compelled to perform power variations on shorter notice (sometimes less than 20 min before the transient), a trend expected to continue and intensify. In such high-stress situations, standard simulation tools are not always convenient to handle because NPP operators often have time to prepare only one scenario and must quickly adapt their actions once the transient begins.

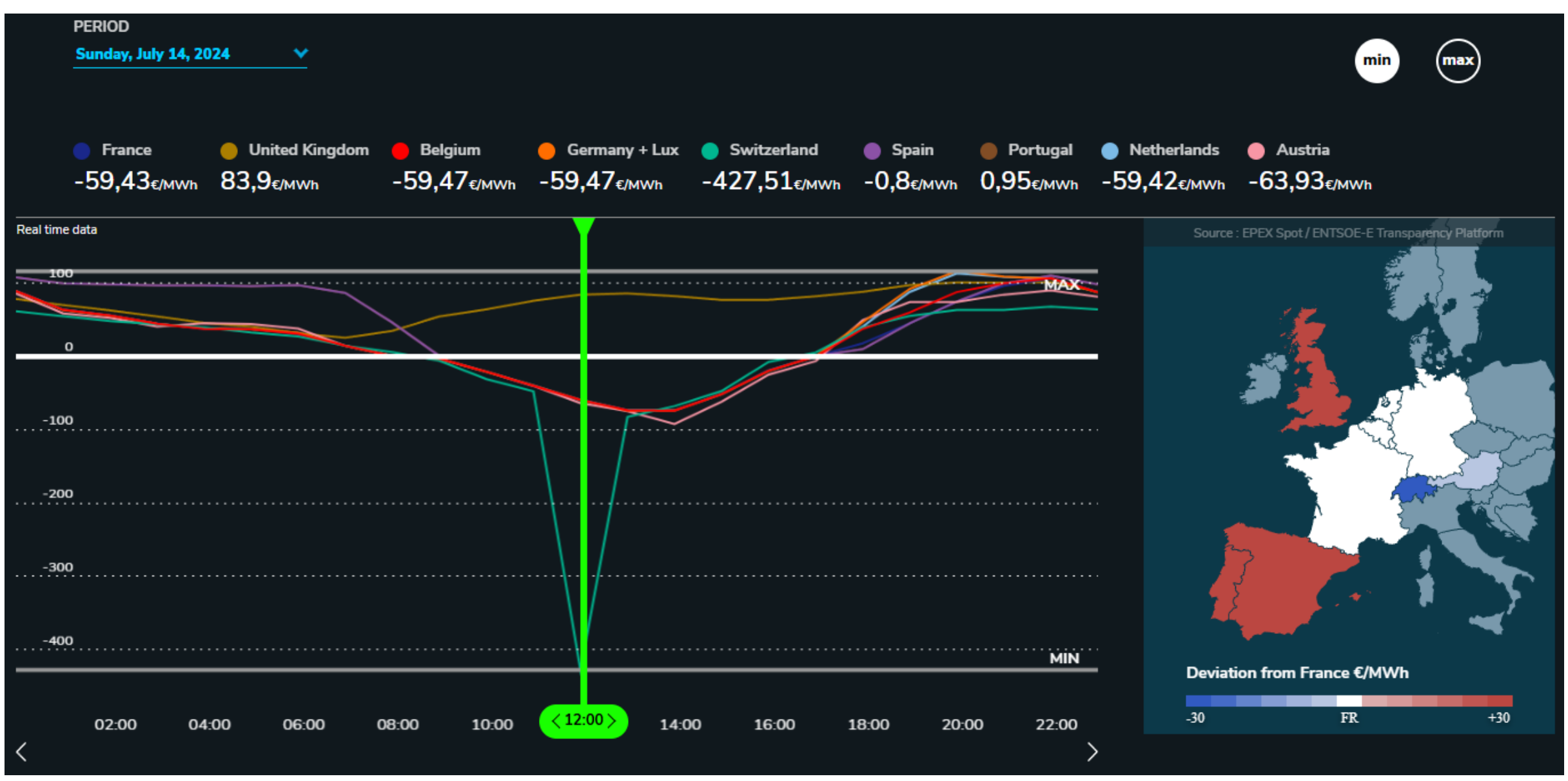

**Figure 1: example of negative electricity prices at noon on July 14, 2024, in Western Europe (screenshot taken from https://www.rte-france.com/en/eco2mix/market-data)**

## 1.3. Overcoming flexibility challenges with the OAPS solution

To address the challenges posed by the evolving energy mix, Framatome has developed an innovative solution for NPPs: the OAPS system [4]. The goal of the OAPS system is to reduce operator burden by automatically providing optimal real-time recommendations during load-following transients (see Figure 2). This GPS-like device improves the flexibility of base-load NPPs and shortens the transient preparation time of already flexible NPPs. Since the role of the OAPS system is limited to showing recommendations to operators, its installation involves only minor modifications to existing equipment and minimal safety classification.

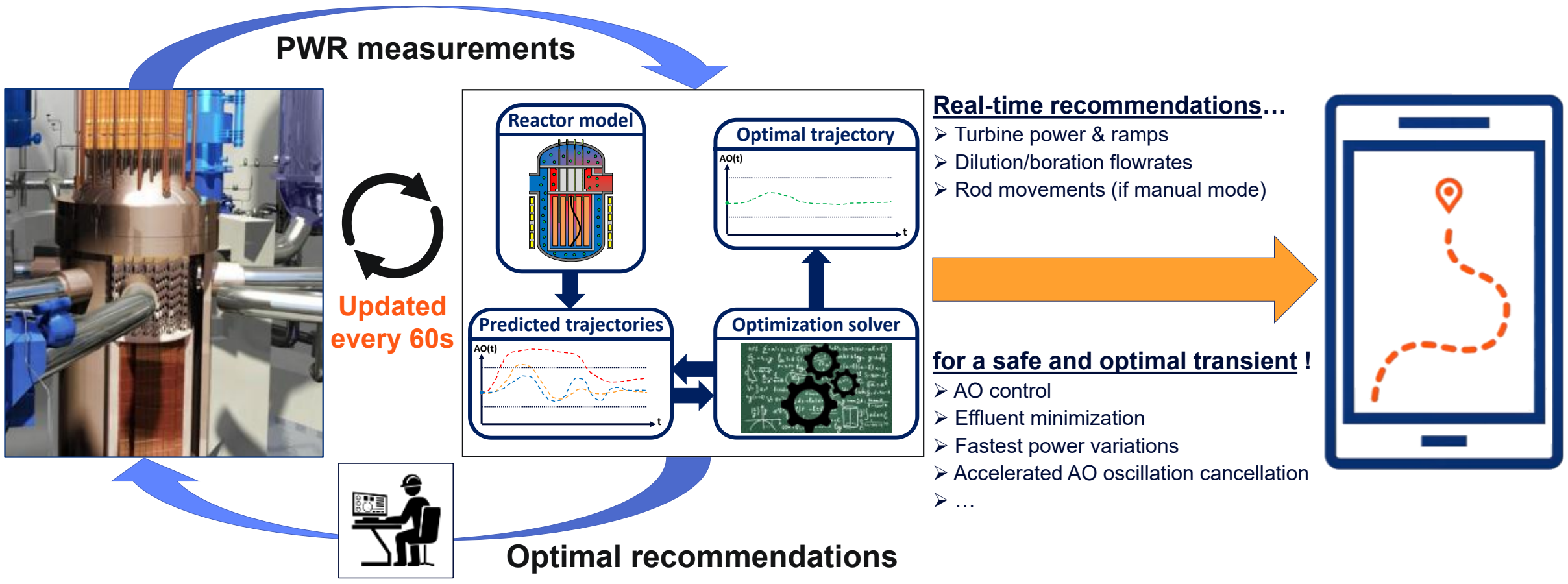

**Figure 2: simplified diagram of the OAPS system principle.**

At the heart of the OAPS system's algorithm is an advanced control technique called model predictive control (MPC). The principle of MPC is to use a model of the plant to predict and optimize its future behavior [5]. More specifically, the model is simulated multiple times with different sequences of control inputs until an optimal solution is found. To keep the plant on track and prevent deviations from the predicted trajectory, new control inputs are continuously recalculated based on the latest measurements. In practice, the optimal sequence of control inputs is computed in real time by solving a constrained optimization problem. Thus, real operational constraints (e.g., LCOs, dilution and boration flowrates limits) and tangible performance metrics (e.g., maximum AO deviation, total effluent volumes) can be directly integrated into the algorithm and easily adapted to user needs. By leveraging the strengths of MPC technology, the OAPS system is able to offer the following unique features:

- **Instant response to last-minute grid requests:** new recommendations can be transmitted within seconds after changing the scheduled power profile of the turbine island on-the-fly.
- **Comprehensive transient optimization:** various customizable strategies can be selected by NPP operators depending on their goals and priorities.
- **Real-time adaptation:** the future recommendations and associated predictions are automatically updated to take account of the latest sensor readings.
- **Continuous monitoring and planning:** the past, current and future evolution of key process variables can always be visualized by NPP operators.

A more detailed description of the OAPS solution is provided in section 3.

## 2. Achieving flexible operation: a complex task

### 2.1. From base-load to load-following operation

When the electrical power produced by the turbine island is constant, e.g., during base-load operation, the thermal power extracted by the steam generators is balanced with the thermal power generated by the reactor core: the NPP is in steady-state and stays around the same operating point. By contrast, when the electrical power produced by the turbine island changes, e.g., during load-following operation, the thermal power extracted by the steam generators no longer matches the thermal power generated by the reactor core: as a result, the average temperature of the primary coolant changes, which in turn affects the axial power distribution of the reactor core through neutronic temperature feedbacks (moderator and Doppler effects). Fortunately, since the fission chain reaction is self-stabilizing by design, the reactor core power will naturally follow the turbine power until they balance out again at the end of the transient. However, if no corrective action is taken, the new operating point may exceed the limiting conditions of operation (LCOs) of the NPP. These limits are defined by upstream engineering studies to achieve optimal performance while complying with safety standards.

Therefore, the main objective of the core control system is to ensure that the average coolant temperature (ACT) and the axial power distribution of the reactor core, or axial offset (AO), respect the LCOs. To achieve this, the core control system can move several control rods within the reactor core or modify the boron concentration of the primary coolant to alter the ACT through neutron absorption (the control rods move rapidly but adversely affect the AO, whereas modifying the boron concentration is much slower but has little to no effect on the AO). The flexibility of the NPP depends mostly on how control rod movements and boron concentration variations are managed.

### 2.2. Should the core control system be redesigned, or the operator augmented?

Three core control systems have been installed on the French nuclear fleet:

- The first one, named Mode A (8 units), was originally designed for base-load operation [6], [7]. In Mode A, a single group of high-absorbing "black" control rods is assigned to ACT regulation. Hence, during power variations, the "black" control rods are automatically inserted in or withdrawn from the reactor core to keep the ACT within its authorized limits. However, this negatively impacts the axial power distribution, as the power of the reactor core changes substantially in the region where the rods are being moved. Thus, to prevent the "black" control rods from moving excessively, the operator must manually adjust the boron concentration of the reactor core by injecting either fresh water (dilution) or borated water (boration) into the primary circuit. Relying solely on dilutions and borations to control the AO during power variations greatly reduces the flexibility of the NPP, as injection flowrates are constrained by the physical limitations of the pumps (typically 10 kg/s for dilution and 3 kg/s for boration). In addition, the effectiveness of dilutions decreases over the fuel burnup cycle because the boron concentration of the primary coolant is gradually reduced to counterbalance uranium depletion (the lower the boron concentration, the less boron is removed for a given volume of injected fresh water). Therefore, in Mode A, the maximum turbine power variation rate is theoretically limited to 1-2 %NP/min at the beginning of the fuel cycle and only 0.1-0.3 %NP/min at 80 % of the fuel cycle. Flexibility is often even more limited in practice, since the operator can struggle to control the AO properly. In fact, unlike control rod movements, the effects of dilutions and borations are delayed by approximately 5 to 15 min, as the injected solutions must flow throughout the chemical and volumetric control circuit before reaching the primary coolant. As a result, AO oscillations may occur if this delay is not correctly anticipated by the operator. Another challenging task for the operator is to mitigate the effect of xenon poisoning on the ACT. More specifically, after a power reduction (resp., increase), the operator must initiate a dilution (resp., boration) to counteract the growth (resp., decay) of xenon. This dilution (resp., boration) must be carefully calculated to avoid triggering the "black" control rods and causing AO disturbances.

- The two other core control systems, named Mode G (48 units) and Mode T (1 unit, fully automatic), were designed for load-following operation [6]-[9]. The key improvement over Mode A is the introduction of a second group of low-absorbing "grey" control rods, which have a lesser impact on the axial power distribution. Having two independently actuated types of control rods enables fast power variations (up to 5 %NP/min). In this configuration, the low-absorbing group can move within the reactor core to control the ACT with minimal AO disturbances, while the high-absorbing group remains in the upper part of the core to slightly adjust the AO – either directly via an automatic control loop in Mode T, or indirectly through manual dilutions and borations in Mode G – if necessary.

Outside of France, the majority of pressurized water reactors, including VVERs [10], are operated in Mode A or its equivalent. One potential solution for improving their maneuverability is to replace the existing core control system with a more flexible option, such as Mode G or Mode T. However, changing the core control system after the commissioning phase requires a complete overhaul of the NPP (e.g., harmonization of monitoring and protection systems, redefinition of the control rod layout, reassessment of operating margins, new safety analyses and licensing, ...). A simpler, more straightforward solution is to maximize the effectiveness of the existing core control system by reinforcing human expertise with optimal assistance.

Consequently, the OAPS system is primarily targeted at base-load NPPs operated in Mode A, or its equivalent, which seek to enhance their flexibility without the need to renovate or replace their existing core control system. Complementary analyses are also being conducted to develop an OAPS system for Mode G and Mode T [11], [12]. In fact, the OAPS system could assist NPP operators in facing new types of emerging transients (e.g., short-duration low-power plateaus) that Mode G and Mode T cannot handle in automatic mode, but which could still be managed manually using the OAPS system.

## 3. Description of the OAPS solution

This section provides an overview of the OAPS system, with a deeper focus on user experience (see Figure 3). Different optimal strategies can be recommended by the OAPS system to suit the needs of NPP operators. The conventional one is to keep the AO close to its reference value to prevent the formation of xenon-induced axial power oscillations. This strategy was validated on two full-scope PWR simulators: first in 2021 with the development team at Framatome, and then in 2023 with two certified NPP operators at EDF DIPDE. Since that time, three new advanced strategies have been implemented in the OAPS system and validated on an intermediate-complexity PWR simulator: 1) fastest power variation rates determination, 2) accelerated AO oscillations cancellation and 3) effluent minimization. In addition, a portable demonstrator with multiple graphical user interfaces (GUIs) has been developed to concretely illustrate how the OAPS system works.

### 3.1. Practical demonstration with the standard axial offset control strategy

To understand how NPP operators could benefit from the OAPS system, a basic load-following scenario is simulated at the beginning of the fuel cycle using the standard AO control strategy. This strategy, which was previously presented in an academic paper [4], is revisited from the perspective of a practitioner interacting with the portable demonstrator and its user-friendly GUIs. For the sake of simplicity, it is assumed that the user perfectly follows the recommendations of the OAPS system, i.e., the recommendations are applied directly and automatically to the NPP.

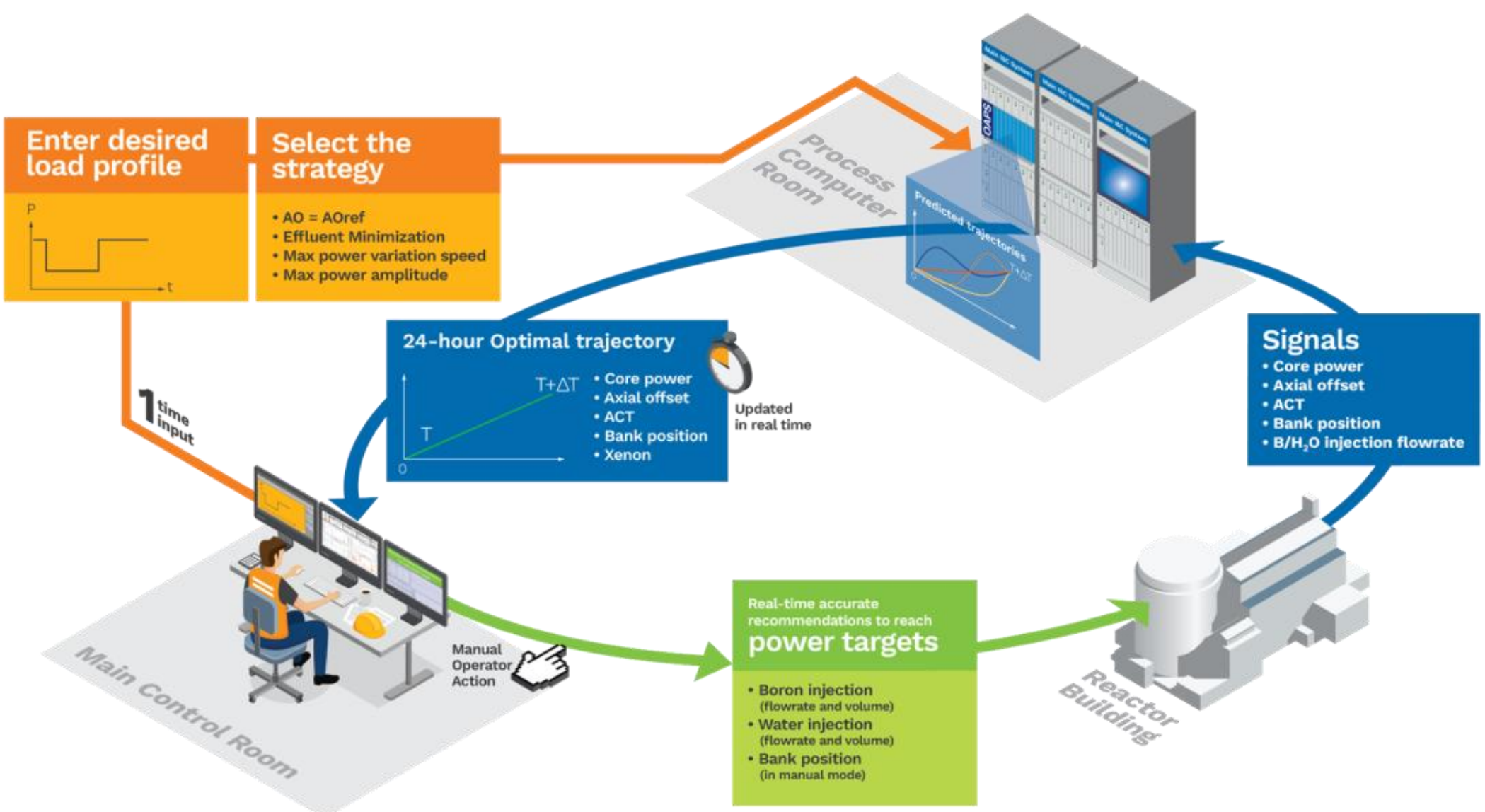

**Figure 3: communication flow between the user, the OAPS system and the NPP.**

The main stages of the demonstration are summarized below:

- First, as shown in Figure 4a, the user only needs to enter the scheduled turbine power profile and specify the reference AO value to configure the transient. Here, two power variations are scheduled at 20 min and 100 min of simulation time, each at a rate of 1 %NP/min.
- Then, as shown in Figure 4b, the OAPS system displays a preview of the entire transient shortly after the computations start. The user can check the evolution of key process variables up to 24 hours in advance.
- Next, as shown in Figure 4c, the recommendations are regularly updated during the simulation using the most recent set of available NPP measurements. Past measurements are represented by solid lines, while future predictions are represented by dashed lines. Note that to avoid overloading the user, the OAPS system prompts NPP operators to change the control input only once every 10 min.
- A zoomed-in view of the AO deviation and dilution/boration flowrate graph, together with the next and current recommendation tables, is presented in Figure 4d. The recommendations that are currently applied to the NPP correspond to the latest ones calculated prior to the transmission date.

Figure 3 synthesizes how the user, the OAPS system, and the NPP interface with each other.

## 3.2. Advanced control strategies

### 3.2.1. Determination of the fastest feasible power variation rates

The first advanced strategy proposed by the OAPS system is to determine the fastest turbine power variation rates that can be performed without exceeding the operational constraints. The goal is to assess the maneuvering capabilities of the NPP and preview the evolution of key reactor core variables (e.g., xenon and boron concentrations or control rods position) over a long-term horizon (typically 24 hours). This strategy is particularly interesting at the end of the fuel burnup cycle, since the rate at which the turbine power can be increased is constrained by dilution efficiency.

To demonstrate, a 24-hour double load-following transient is simulated at 80 % of the fuel cycle using the fastest power variation rates determination strategy. As observed in Figure 5a and Figure 5b, the OAPS system can help reduce the time required to return to full power by gradually increasing the turbine power variation rate as the xenon concentration decreases. The AO deviation, shown in Figure 5d, stays within the authorized limits (green dotted lines at $\pm 5$ %) throughout the simulation, even for power reduction rates greater than 2 %NP/min.

### 3.2.2. Accelerated cancellation of axial power oscillations

In certain specific situations, maintaining strict AO control may become secondary or even impossible.

For example, maximum injection flowrates or control rod position limits can be reached during an emergency power variation, resulting in a temporary loss of AO control. Periodic core testing can also lead to intentional AO deviations. To prevent xenon-induced axial power oscillations from increasing after these unusual transients, NPP operators must bring the reactor core back to equilibrium as soon as possible. The simplest approach to cancel axial power oscillations is to follow the conventional AO control strategy, i.e., to maintain the AO on its reference value until xenon equilibrium is achieved. However, this is not always feasible and tends to be relatively slow. A simple yet faster strategy proposed by the OAPS system is to purposely increase the AO deviation (see Figure 6a) to steer the axial iodine imbalance toward its equilibrium value (see Figure 6b). This strategy is challenging for NPP operators to follow without assistance, especially since the iodine and xenon concentrations cannot be measured.

The accelerated AO oscillations cancellation strategy (blue lines) is compared to the standard AO control strategy (red lines) on a 48-hour transient simulated at the beginning of the fuel cycle. In both scenarios, a xenon-induced AO oscillation, generated by a power reduction from 100 to 50 %NP, builds up during the first 8 hours of simulation time. After that, the OAPS system is activated to cancel the oscillation. The accelerated strategy works properly since the iodine and xenon imbalances converge approximately 50 % faster than with the conventional strategy (see Figure 6b and Figure 6c). The OAPS system can carry out this complex strategy effortlessly thanks to its ability to predict both quantities.

### 3.2.3. Minimization of water and boron effluents

Whenever fresh or borated water is injected into the primary circuit, an equivalent amount of primary coolant must be removed to maintain mass balance. These effluents need to be stored and processed before being either reused or discharged into the environment. Thus, minimizing effluent production is an effective way of optimizing costs while reducing environmental impact. The strategy proposed by the OAPS system to save effluents over a daily load-following cycle is to relax operational constraints at low power levels, within acceptable ranges, and take advantage of xenon poisoning.

The effluent minimization strategy (blue lines) is compared to the standard AO control strategy (red lines) on a 24-hour load-following transient simulated at the beginning of the fuel cycle. As seen in Figure 7a, the operational AO constraints (purple dashed lines) have been significantly extended at 50 %NP. Hence, during the low-power plateau, the AO can freely evolve in response to xenon-induced control rod movements. Thus, fewer boron concentration variations are needed, as shown in Figure 7c, because the goal is not to control the AO as precisely as in the standard strategy. In fact, at the end of the transient, the cumulative mass of borated and fresh water injected is decreased by about 25 %, as can be observed in Figure 7b. By adapting operational constraints and anticipating xenon evolution, the OAPS system can help NPP operators optimize effluent production in a safe and efficient way.

## 4. Conclusion

In this paper, an innovative real-time predictive system for flexible PWR operation has been presented. This new device, the OAPS system, can enhance the flexibility of base-load NPP and shorten the transient preparation time of already flexible NPPs. By taking into account real operational constraints and tangible performance metrics, the OAPS system can offer various optimal control strategies to NPP operators: 1) AO control, 2) fastest power variation rates determination, 3) accelerated AO oscillations cancellation and 4) effluent minimization. In addition, thanks to its predictive capabilities, the OAPS system can help NPP operators monitor and forecast the evolution of key process variables over a long-time horizon. An interactive portable demonstrator has also been developed to showcase the unique features of the OAPS system in a simple and intuitive manner.

## Appendix A: standard AO control strategy in the portable OAPS system demonstrator

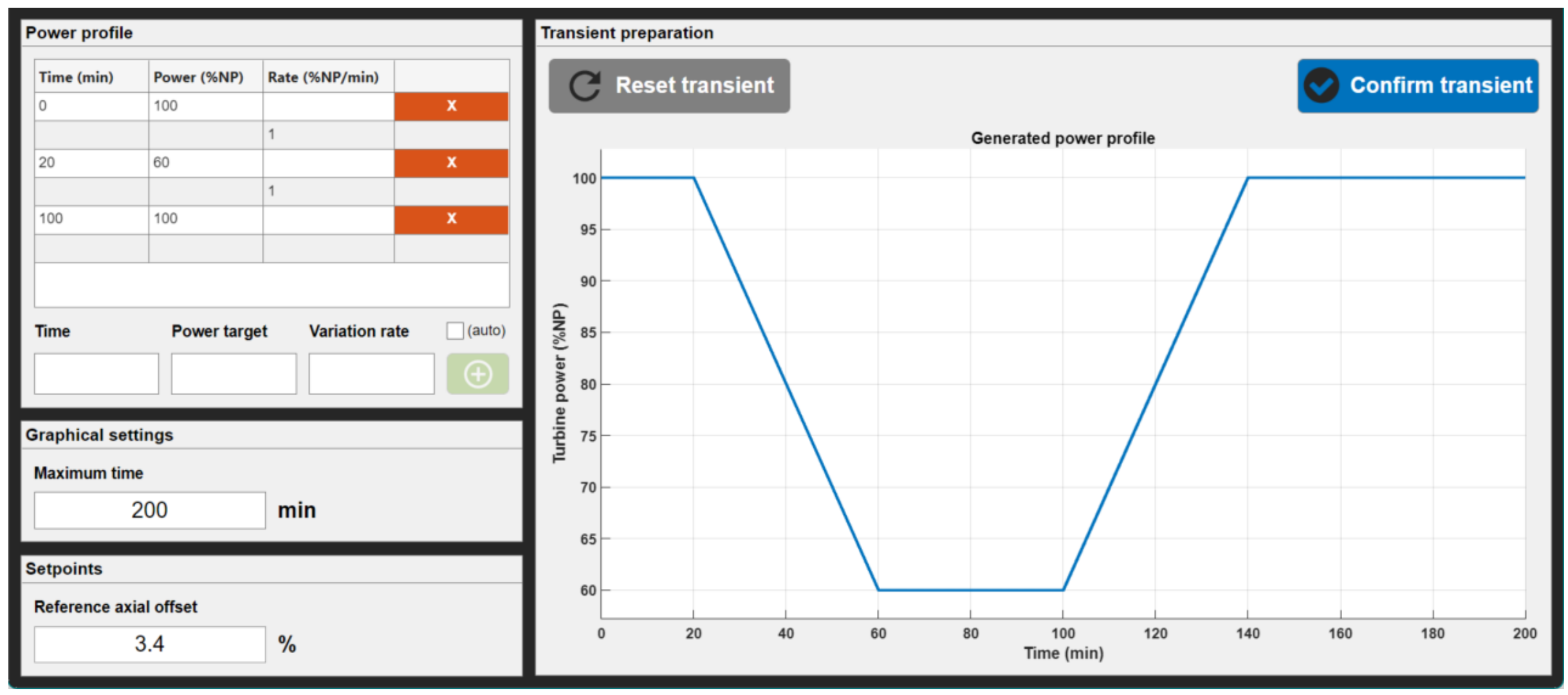

**Figure 4a: transient preparation interface.**

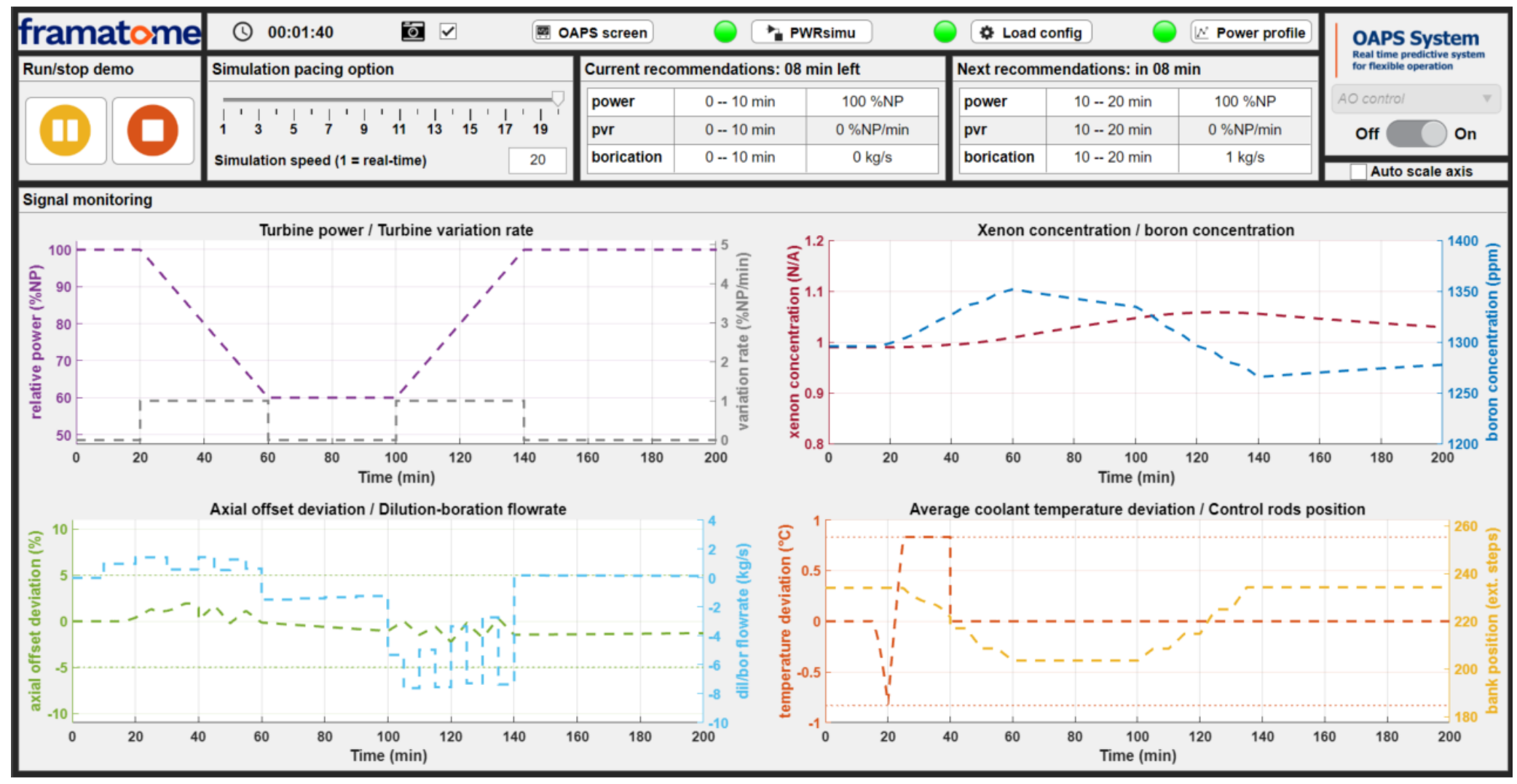

**Figure 4b: first set of OAPS system predictions.**

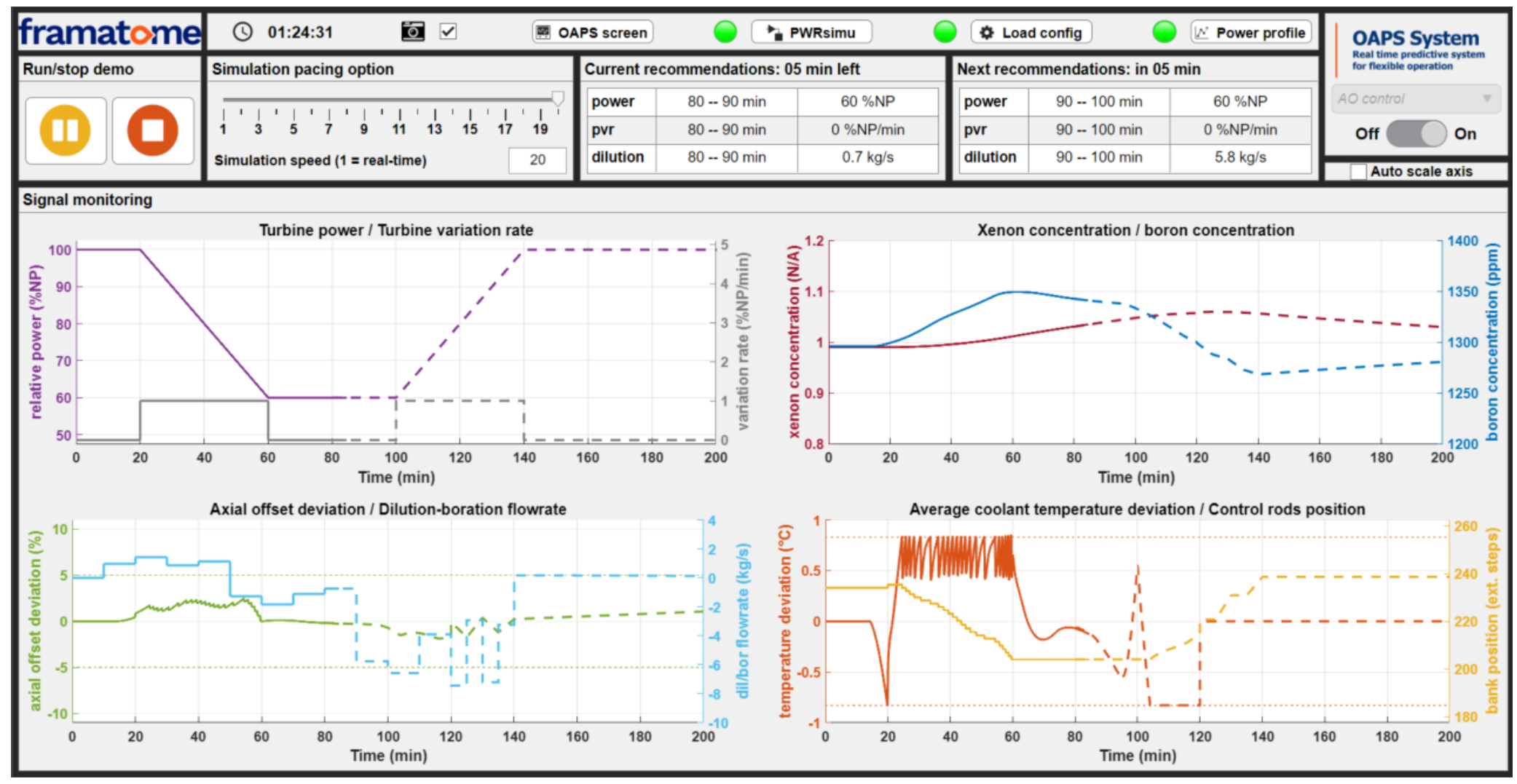

Figure 4c: NPP measurements (solid) and OAPS predictions (dashed).

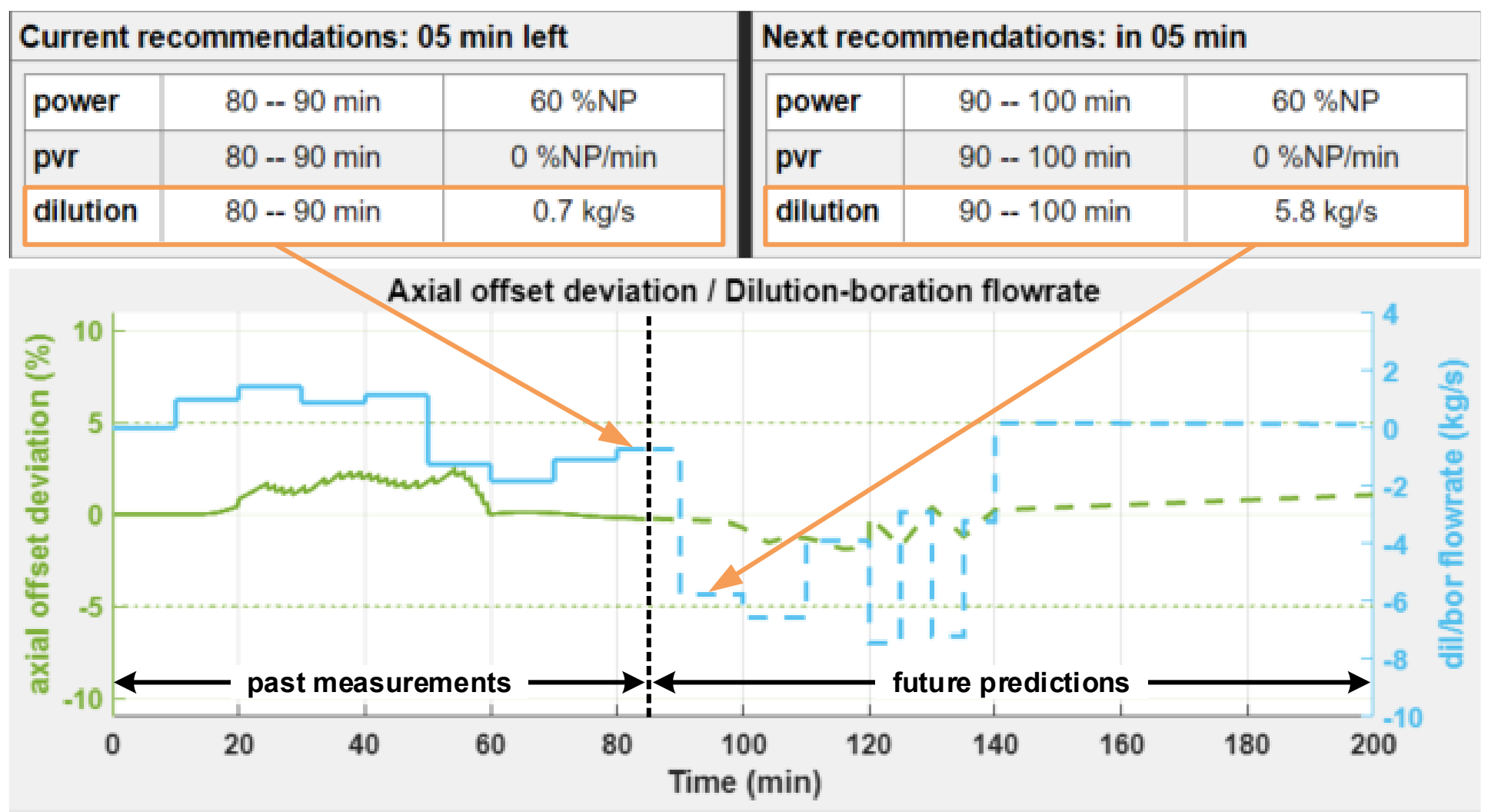

Figure 4d: focus on the current and next dilution/boration flowrate recommendations.

## Appendix B: fastest power variation rates determination strategy

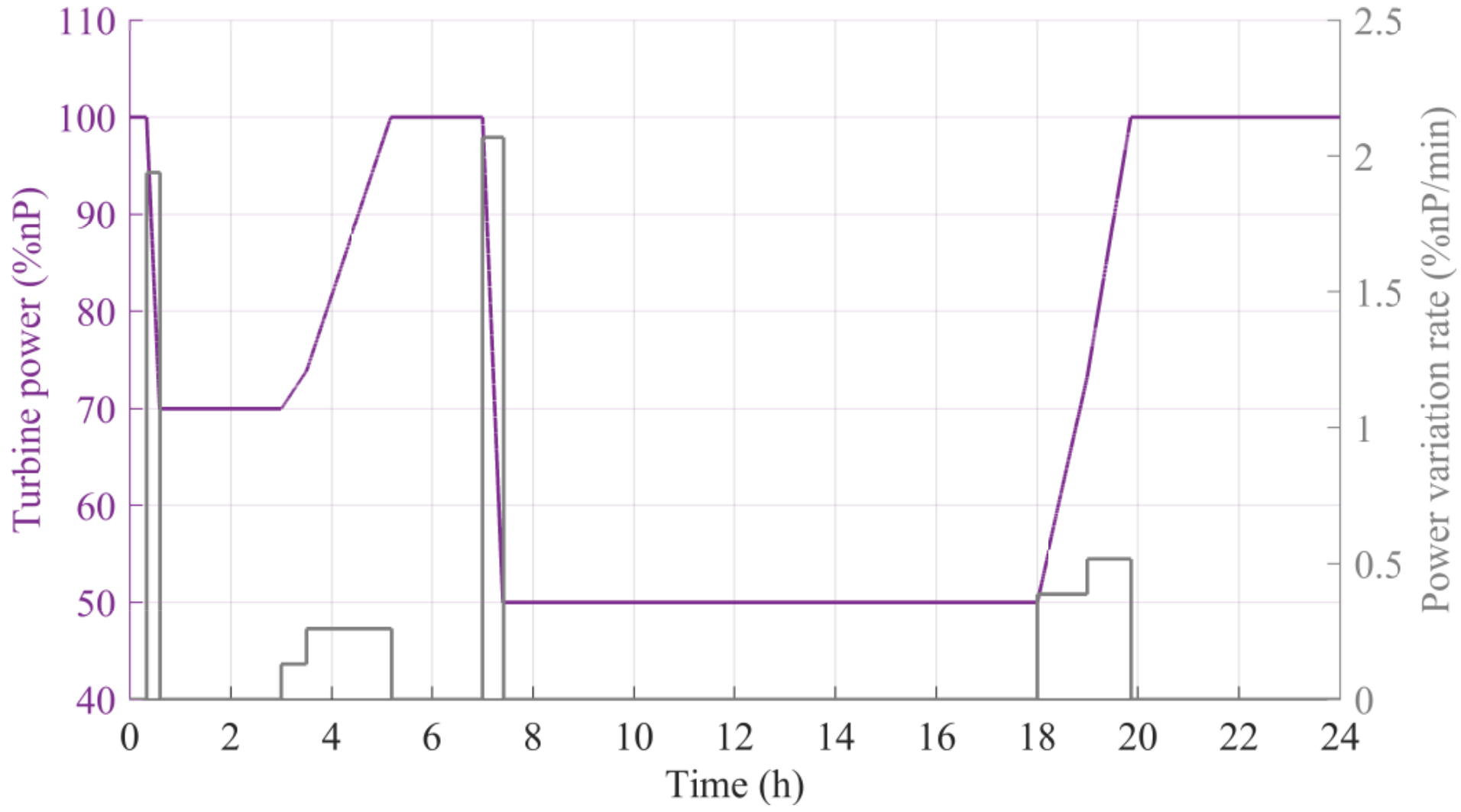

Figure 5a: turbine power (purple) and power variation rate (grey).

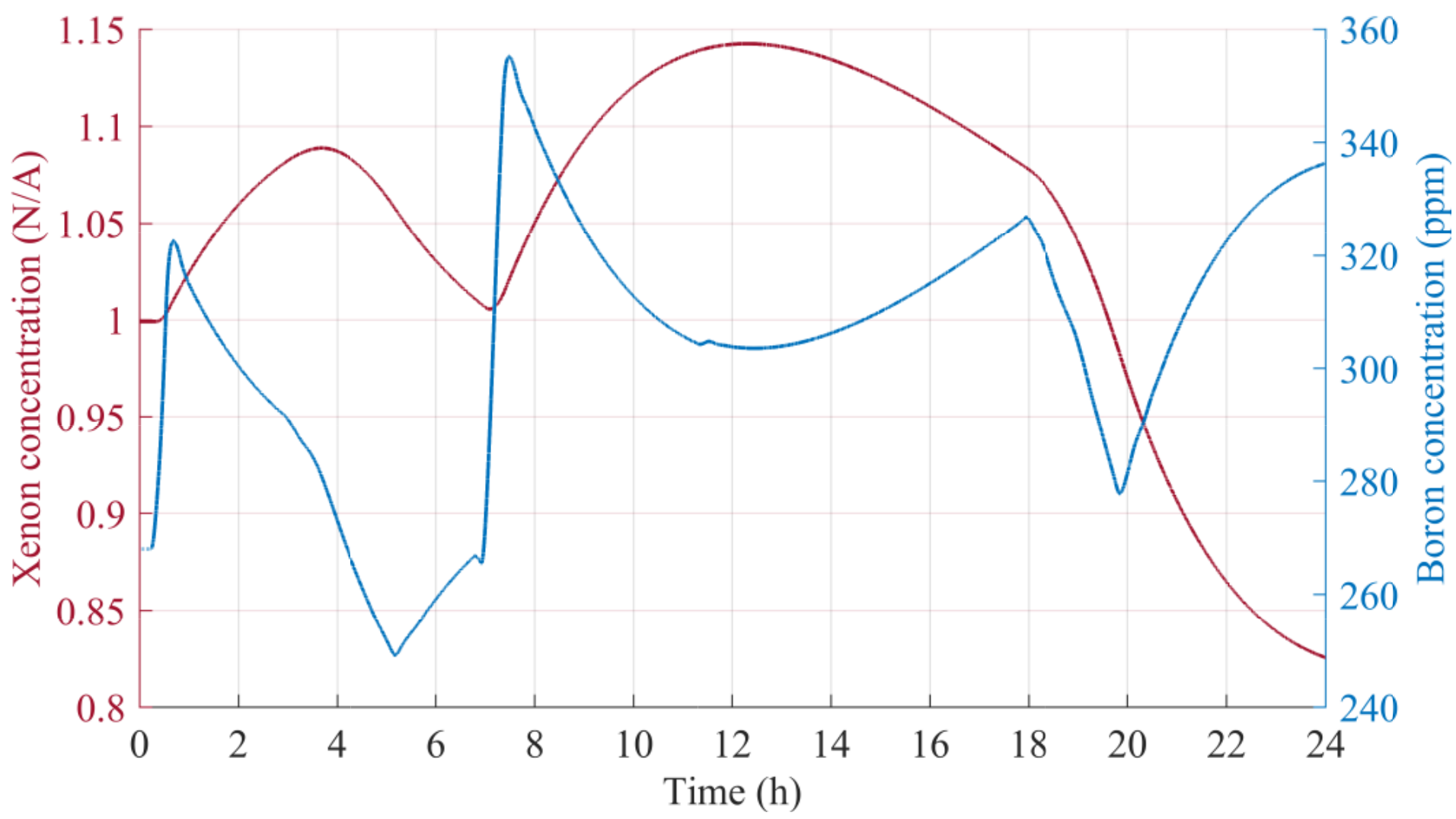

**Figure 5b: xenon (burgundy) and boron (blue) concentrations.**

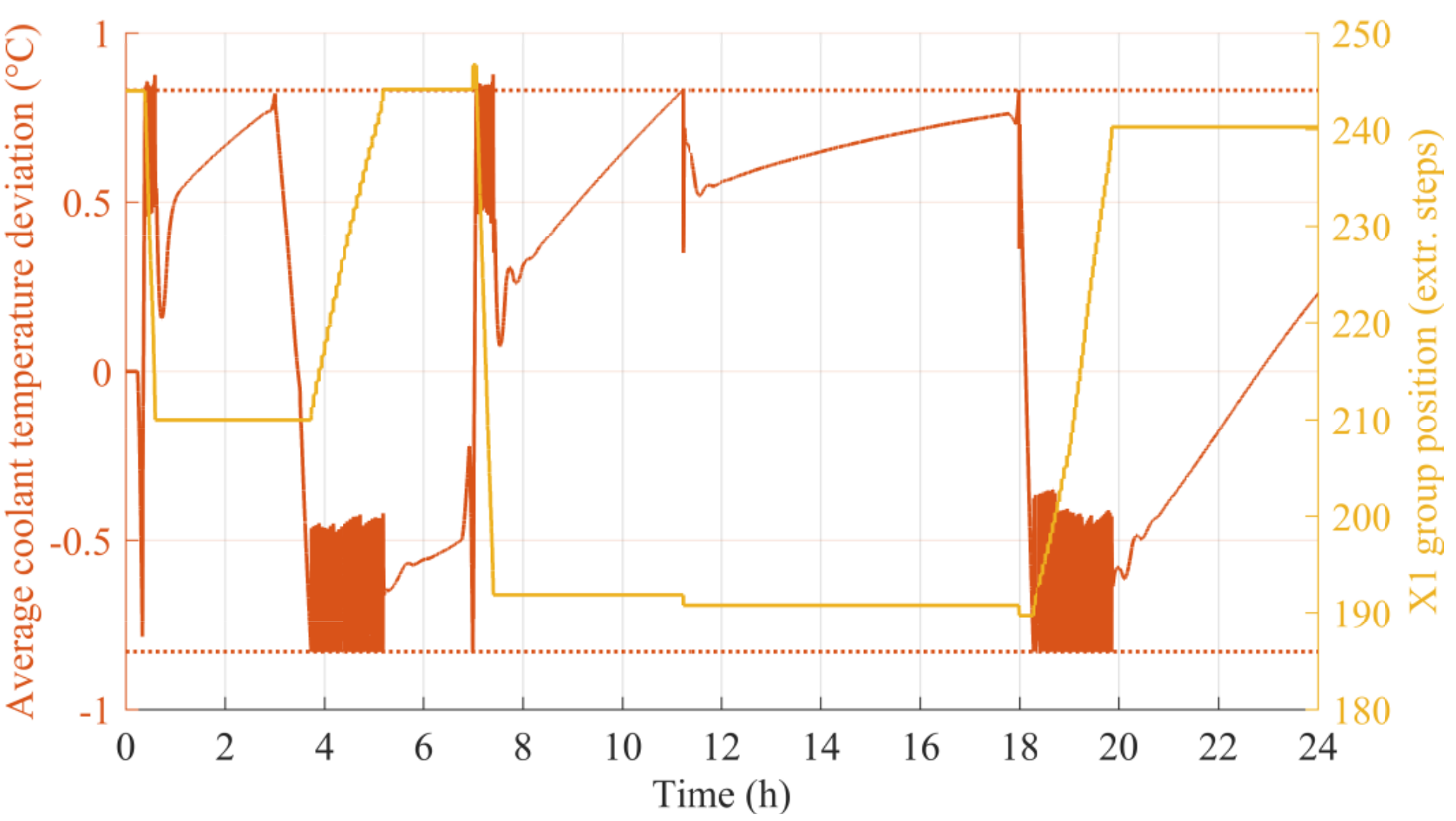

**Figure 5c: ACT deviation (red) and X1 bank position (yellow).**

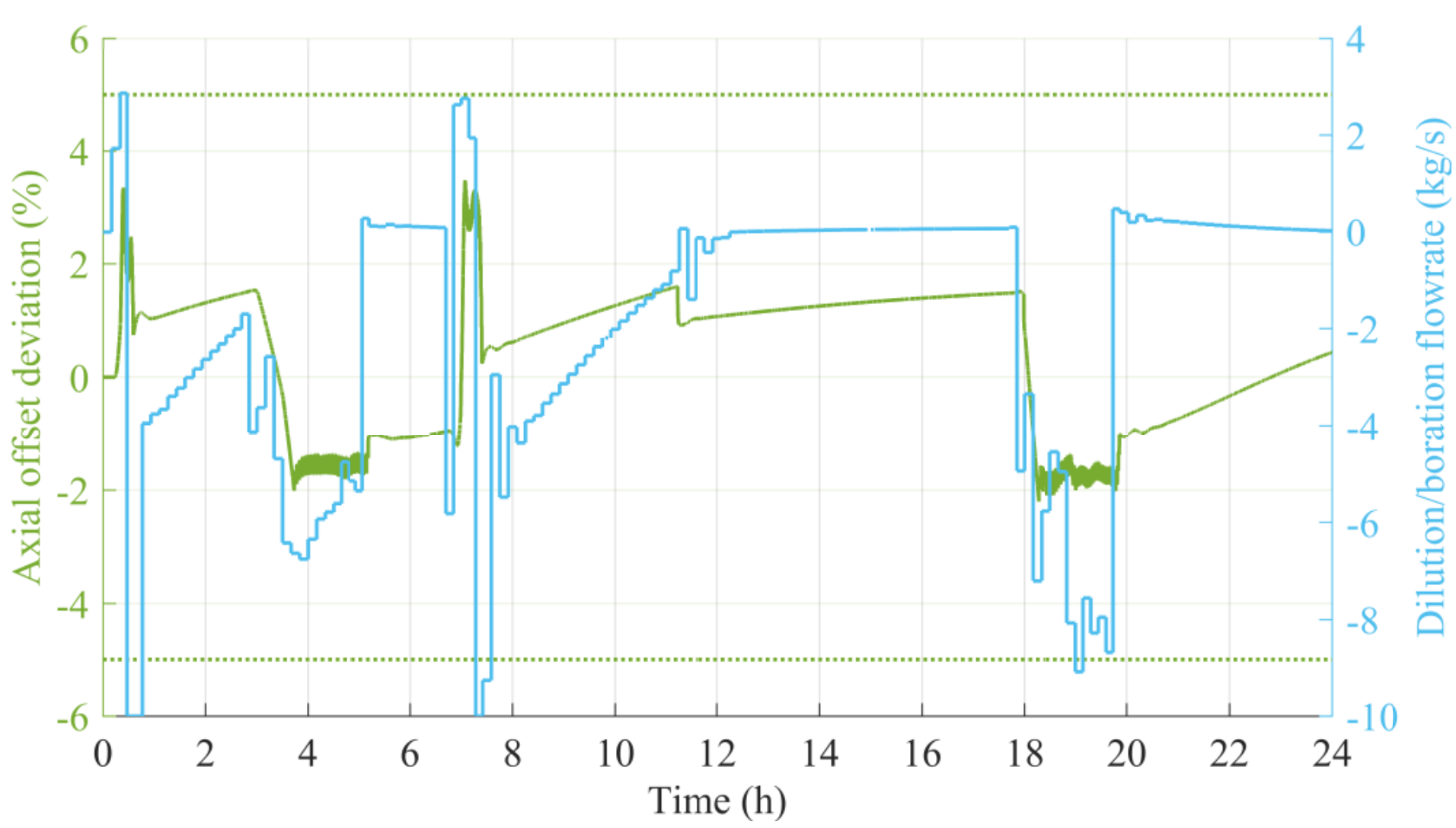

**Figure 5d: AO deviation (green) and dilution/boration flowrate (cyan).**

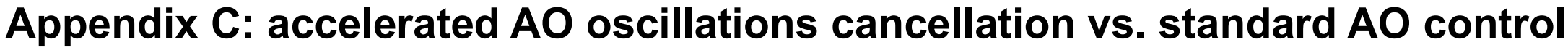

## Appendix C: accelerated AO oscillations cancellation vs. standard AO control

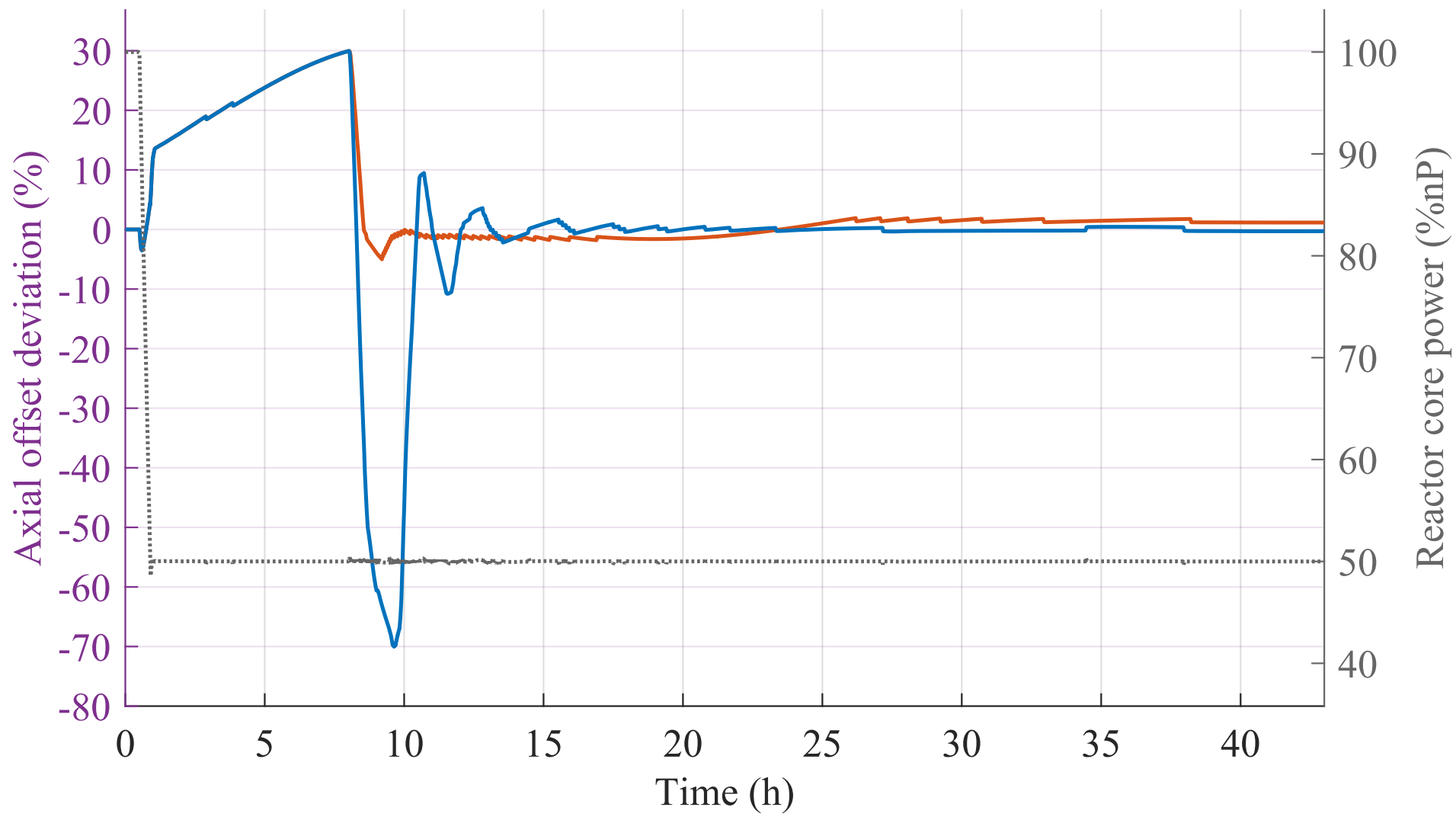

**Figure 6a: AO deviation simulated with the accelerated (blue) and standard (red) strategies.**

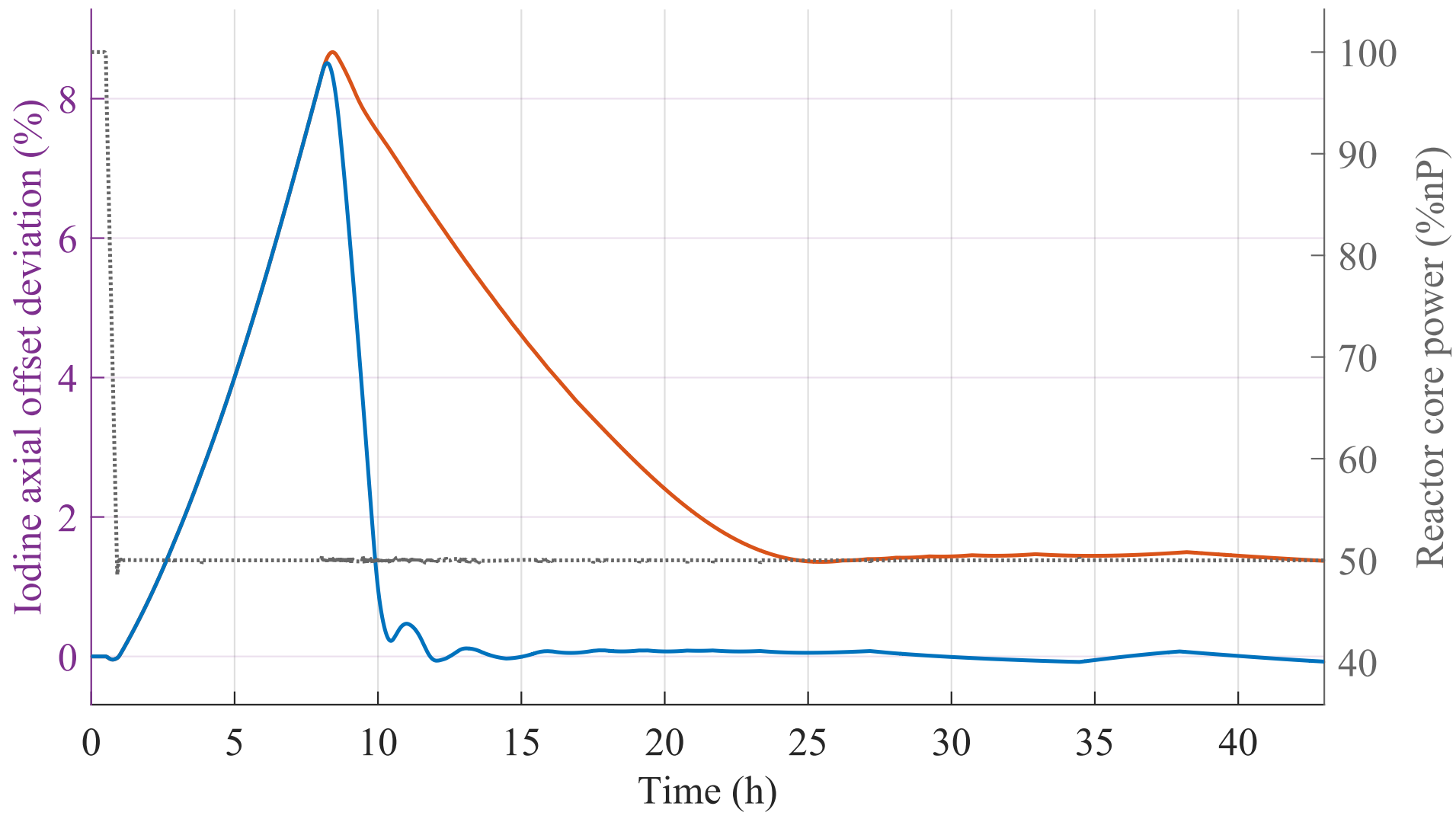

**Figure 6b: iodine axial offset deviation simulated with the accelerated (blue) and standard (red) strategies.**

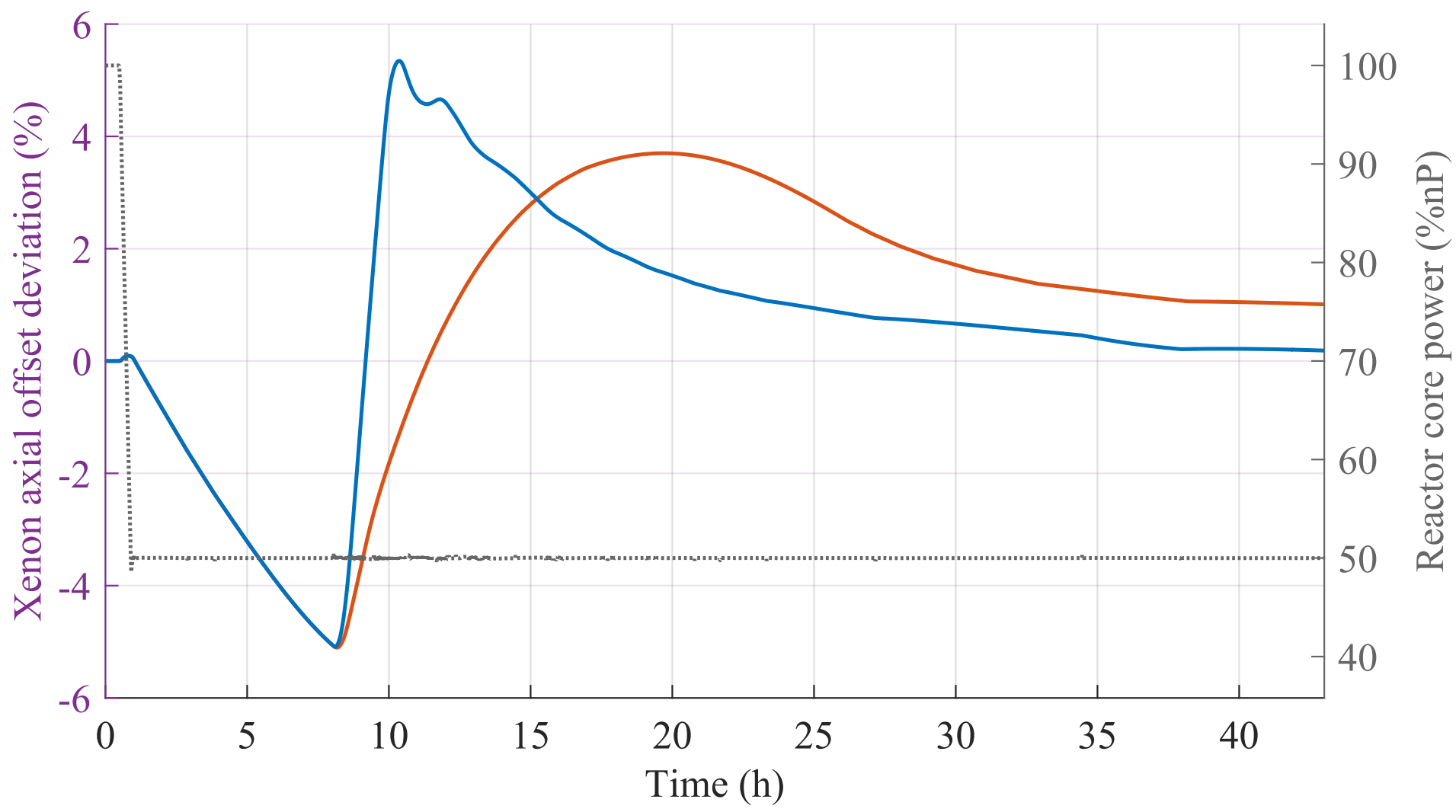

**Figure 6c: xenon axial offset deviation simulated with the accelerated (blue) and standard (red) strategies.**

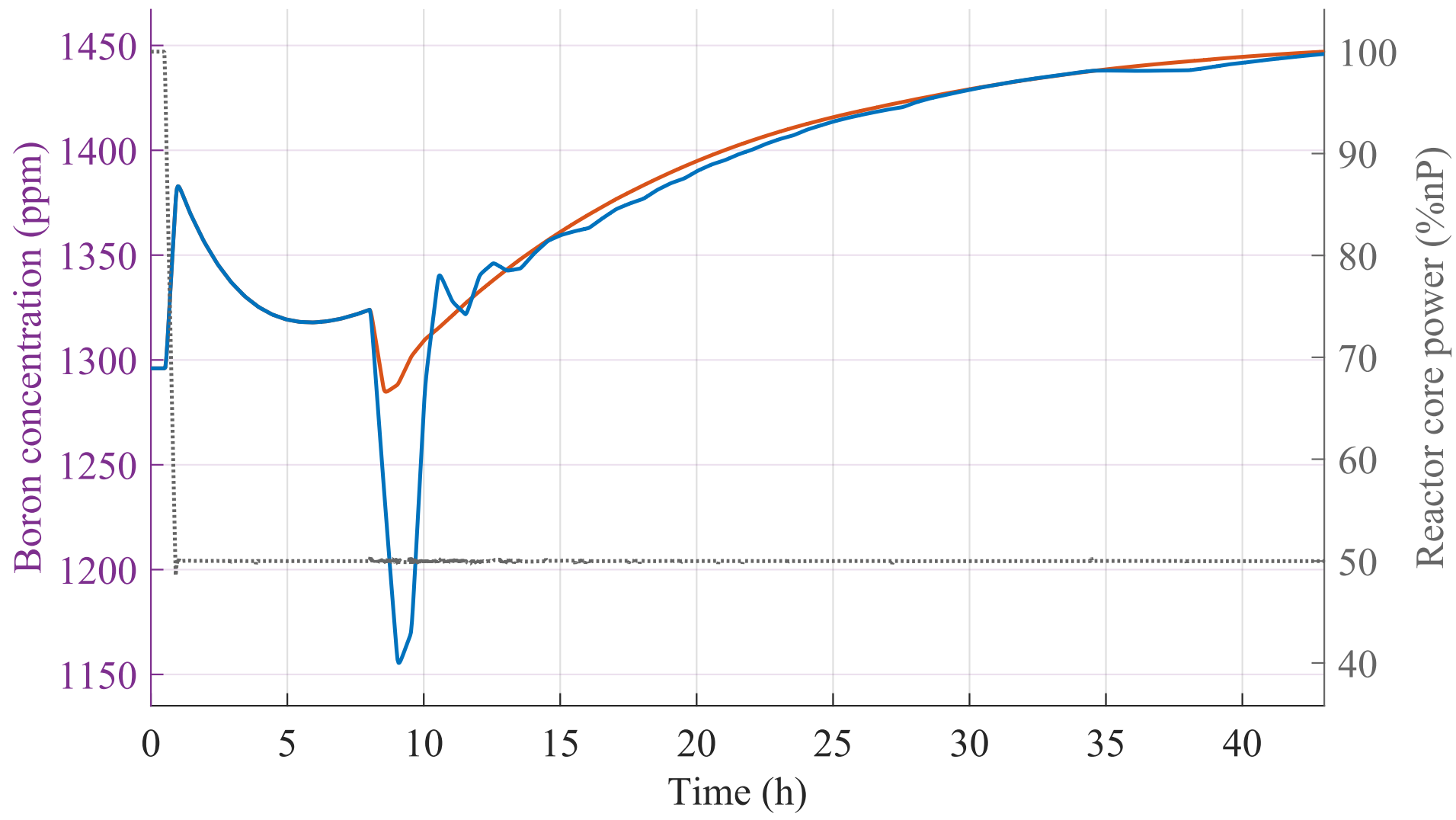

**Figure 6d: boron concentration simulated with the accelerated (blue) and standard (red) strategies.**

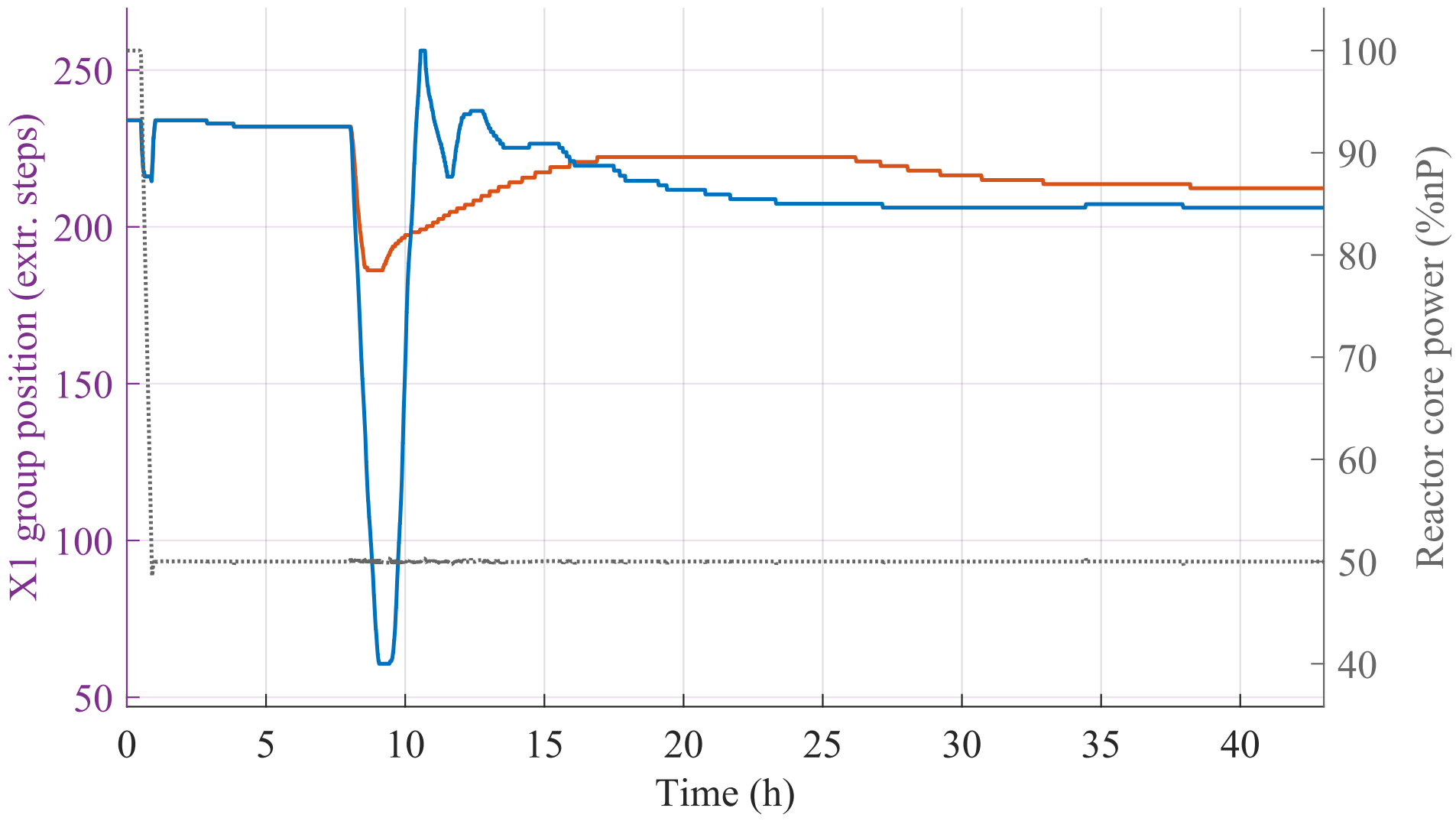

**Figure 6e: X1 bank position simulated with the accelerated (blue) and standard (red) strategies.**

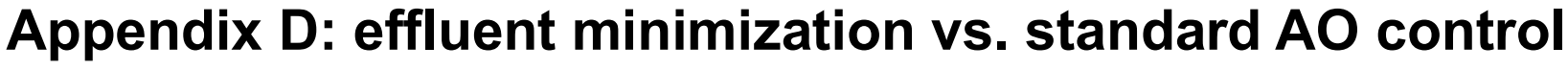

## Appendix D: effluent minimization vs. standard AO control

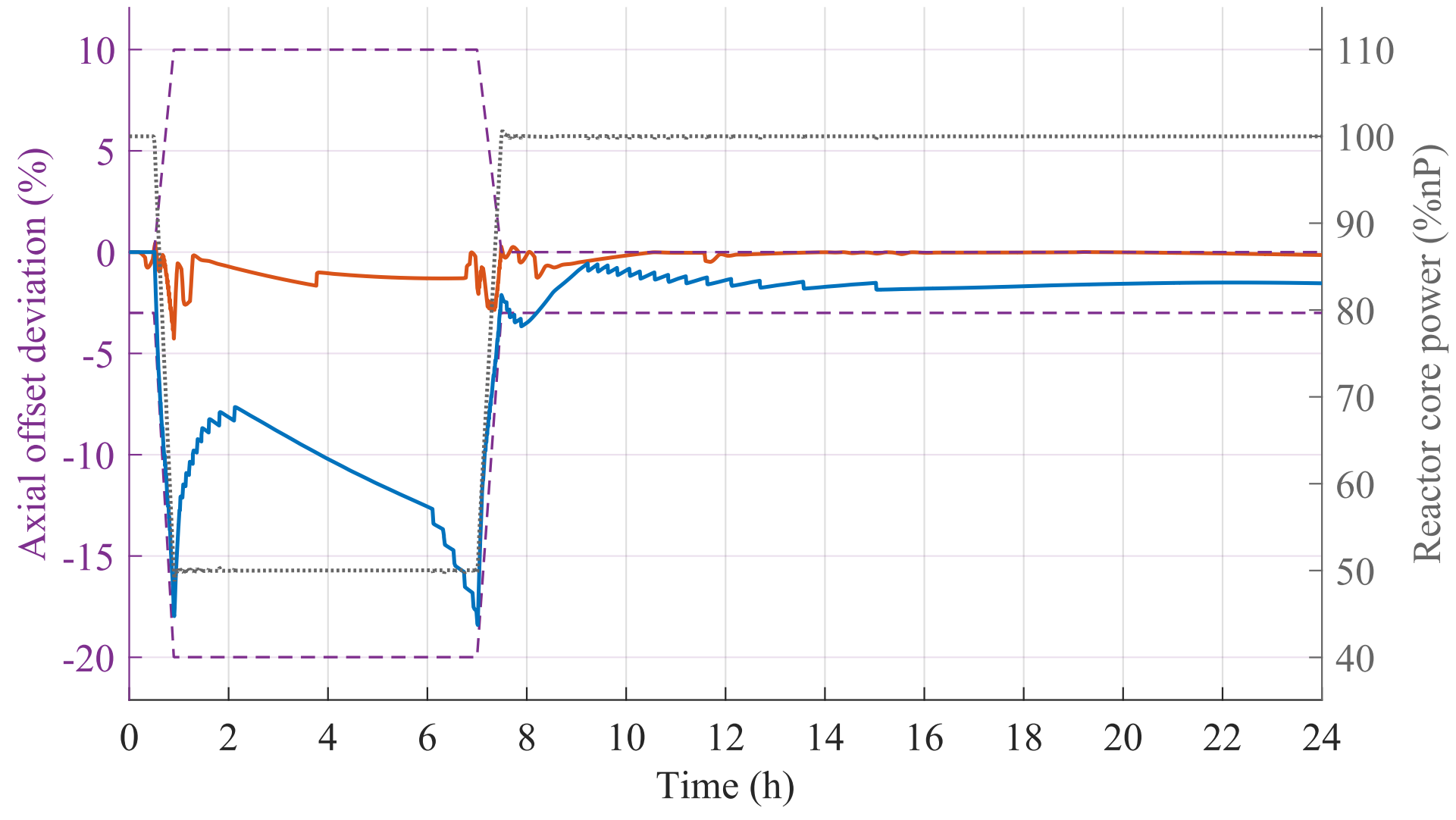

**Figure 7a: AO deviation simulated with the minimization (blue) and standard (red) strategies.**

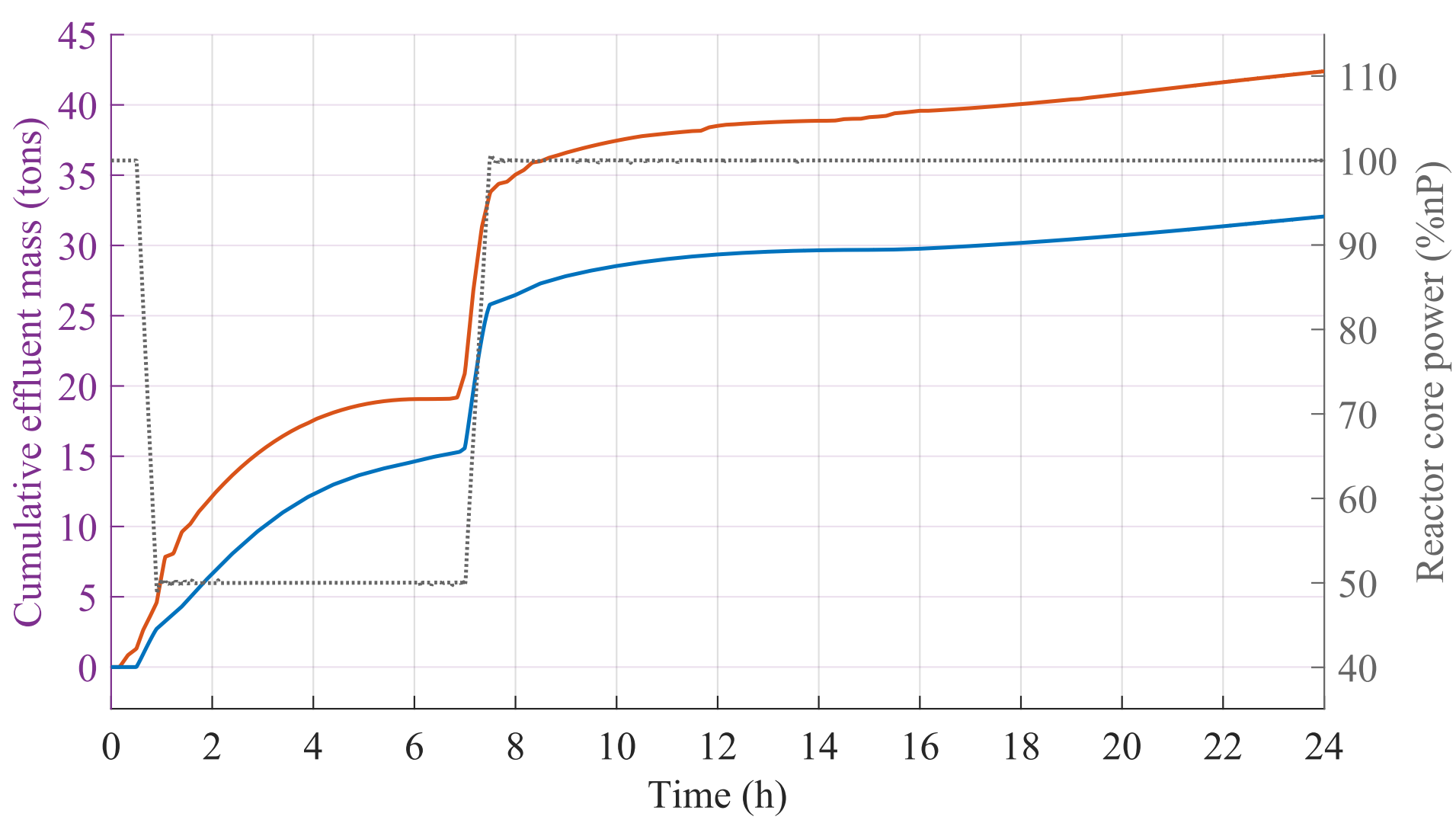

**Figure 7b: cumulative effluent mass simulated with the minimization (blue) and standard (red) strategies.**

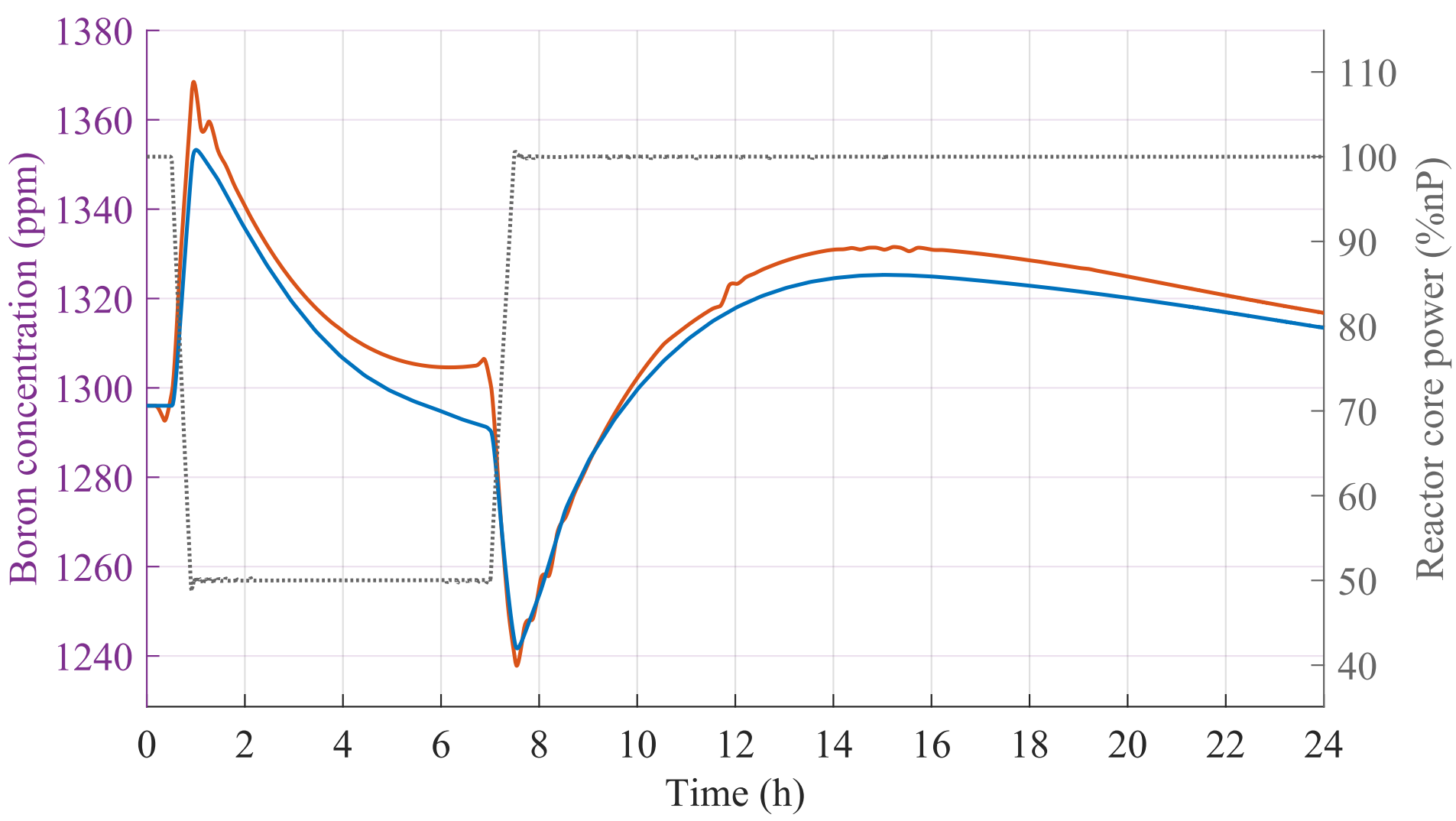

**Figure 7c: boron concentration simulated with the minimization (blue) and standard (red) strategies.**

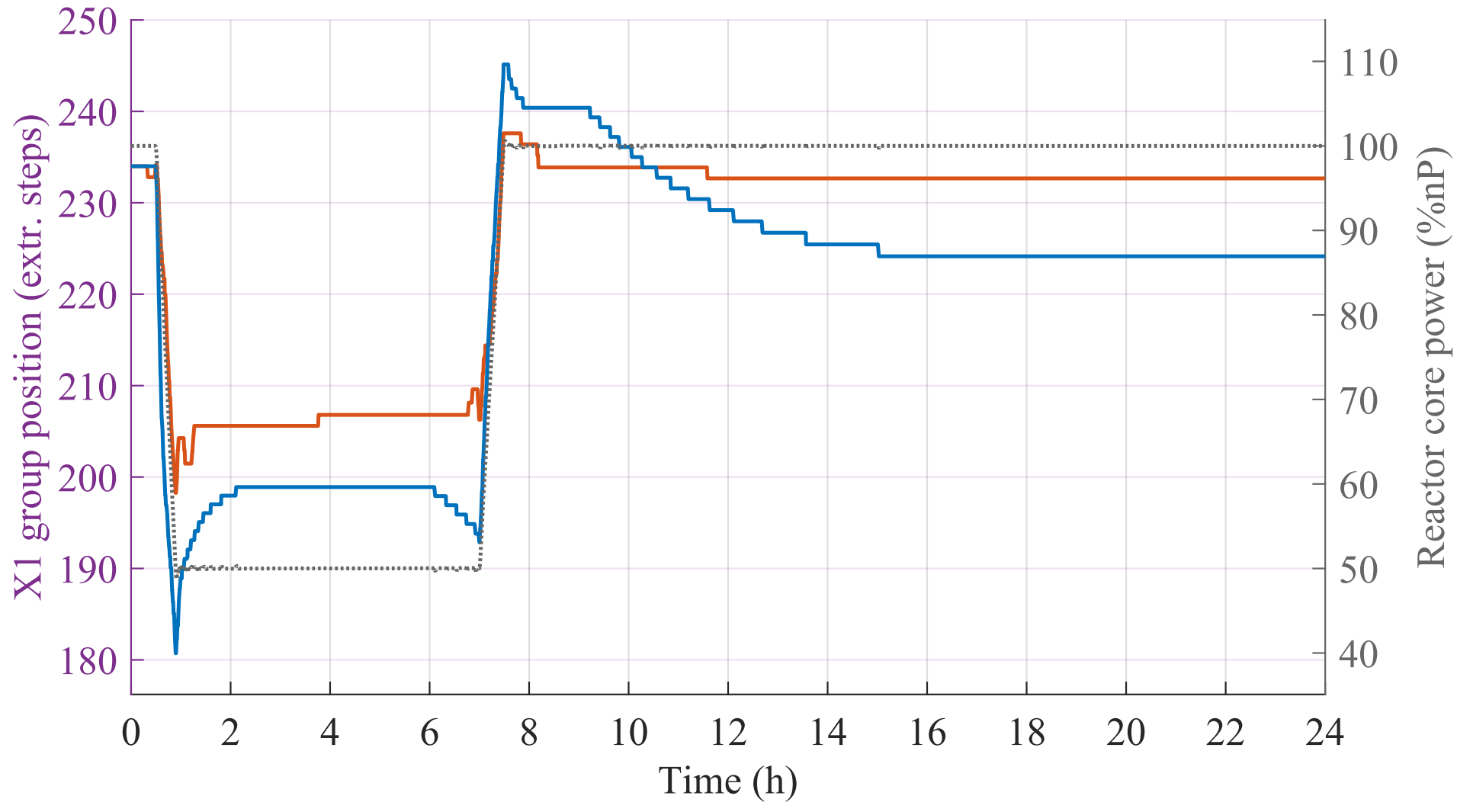

**Figure 7d: X1 bank position simulated with the minimization (blue) and standard (red) strategies.**